\pdfoutput=1

\documentclass[12pt, a4paper]{article}

\usepackage{notes2bib}
\usepackage{amsmath}
\usepackage{amsfonts}
\usepackage{amssymb}
\usepackage{bbm}
\usepackage{verbatim}
\usepackage{dsfont}
\usepackage{booktabs}
\usepackage{slashed}
\usepackage{bm}
\usepackage{enumitem}

\usepackage{epsfig}
\usepackage[table]{xcolor}
\usepackage{graphicx}
\usepackage[]{caption}
\usepackage[listofformat=empty,subrefformat=empty]{subfig}

\usepackage{float}
\usepackage{overpic}

\usepackage{enumerate}
\usepackage{hhline}
\usepackage{multirow}
\usepackage[hidelinks]{hyperref}

\usepackage{cite}
\usepackage{a4wide}
\usepackage{bbold}

\usepackage{geometry}
\renewcommand{\baselinestretch}{1.2}

\newcommand{\Dh}{\Delta_h}

\definecolor{darkgreen}{HTML}{3C8031}

\begin{document}

\thispagestyle{empty}

\begin{flushright}
DESY 26-121 
\end{flushright}
\vskip .8 cm
\begin{center}
  {\Large \bf Probability measures for de
    Sitter JT gravity}
 \\[12pt]

\bigskip 
{
{\bf{Wilfried Buchm\"uller}\footnote{E-mail:
    wilfried.buchmueller@desy.de}}
and
{\bf{Alexander Westphal}\footnote{E-mail:
    alexander.westphal@desy.de}}
}
\bigskip\\[0pt]
\vspace{0.23cm}
{\it Deutsches Elektronen-Synchrotron DESY, Notkestr. 85,
  22607 Hamburg, Germany}
\\[20pt] 
\bigskip
\end{center}

\date{\today}

\begin{abstract}
\noindent
Jackiw-Teitelboim (JT) gravity with positive cosmological constant
is a toy model to study the emergence of expanding de Sitter universes from
a quantum state. Its density matrix is a superposition of  double-trumpet
amplitudes which are solutions of a Wheeler-DeWitt equation. We
compare predictions of such mixed states for classical expanding
universes with the corresponding predictions of no-boundary wavefunctions.
Double-trumpet amplitudes consistent with holography, whose
semiclassical limit matches the result from Schwarzian boundary fluctuations,
lead to a flat probability distribution for the size of the universe,
consistent with models of inflation. This is in contrast to the
suppression of large universes predicted by no-boundary wave functions.
\end{abstract}

\newpage 
\setcounter{page}{2}
\setcounter{footnote}{0}
{\renewcommand{\baselinestretch}{1}\tableofcontents}

\section{Introduction}
\label{sec:introduction}

Understanding the emergence of classical universes from a quantum
state remains an unsolved problem in quantum gravity and cosmology.
Important advances include the proposal of Euclidean \cite{Hartle:1983ai} and
Lorentzian \cite{Linde:1983mx,Vilenkin:1984wp} ``wavefunctions of the
universe'' (for a review, see \cite{Lehners:2023yrj}) and their
  cosmological applications in the semiclassical regime (for early
reviews, see \cite{Halliwell:1992cj,Vilenkin:1993ym}).
However, the goal to derive testable predictions, consistent
with the observed cosmic microwave background,  remains challenging
(see, for example, \cite{Hartle:2008ng,Maldacena:2024uhs}).

In recent years, important progress has been made in two-dimensional (2d)
models of quantum gravity. A particularly interesting example 
is Jackiw-Teitelboim (JT) gravity \cite{Teitelboim:1983ux,Jackiw:1984je}. The
model is exactly solvable \cite{Henneaux:1985nw,Louis-Martinez:1993bge} and,
like all 2d dilaton-gravity theories, its minisuperspace version
already contains all physical information
\cite{Louis-Martinez:1993bge}. Over the past years, JT gravity has been
studied in detail in anti-de Sitter ($\text{AdS}$) space (for a
review, see, for example, \cite{Mertens:2022irh}),
and, more recently, also in de Sitter ($\text{dS}$) JT gravity at large field values 
\cite{Maldacena:2019cbz,Cotler:2019nbi}. This is achieved
by reducing the path integral for the wavefunction to
a path integral over the Schwarzian degrees of freedom
of a boundary curve, as in $\text{AdS}_2$ gravity
\cite{Stanford:2017thb}. Correspondingly, exact solutions to the
Wheeler-DeWitt (WDW) equation with Schwarzian asymptotic behaviour have been
analysed in \cite{Buchmuller:2024ksd}. dS JT gravity is also an
interesting toy model for cosmological questions (see, for example,
\cite{Anninos:2024iwf,Alexandre:2025rgx,Dey:2026lcu}).

The action of JT gravity is linear in the dilaton field. Hence, the
dilaton can be integrated out, and only bulk metrics with constant
curvature contribute to the gravitational path integral.
This implies that gravitational amplitudes can be composed of
``trumpet amplitudes'' for the transition between a geodesic surface and
a second surface with Dirichlet boundary conditions
\cite{Saad:2019lba}. In particular, an amplitude with two surfaces is
given by a double-trumpet amplitude. This is true for Euclidean
as well as Lorentzian geometries. At infinite cutoff the trumpet amplitudes for
$\text{EAdS}_2$ and $\text{dS}_2$ have been computed in \cite{Saad:2019lba}  
and \cite{Cotler:2019nbi}, respectively. Summing up an infinite series of 
extrinsic curvature terms of the boundary curve, trumpet and
no-boundary wavefunctions have been computed in
\cite{Iliesiu:2020zld} for $\text{EAdS}_2$. Recently, these wavefunctions
have also been obtained from a bulk path integral
\cite{Griguolo:2025kpi}. The two methods lead to consistent results.
The wavefunctions satisfy WDW equations with non-flat operator ordering.

A major goal in quantum cosmology is a prediction for the size of the
universe. The corresponding probability distributions are mostly based
on the Klein-Gordon (KG) current that is conserved 
for solutions of the WDW equation \cite{Vilenkin:1988yd}. For recent
examples, see \cite{Hartle:2008ng,Maldacena:2019cbz,Blommaert:2025rgw}.
Typically, these probability distributions fall off at large universe
size. This is welcome in view of the normalizability of the
wavefunction. However, it leads to the problem that the probability
for large universes is suppressed, which is inconsistent with
inflation \cite{Hartle:2008ng,Maldacena:2024uhs}.

A different perspective on this problem is provided by considering a
conditional density matrix for the computation of a probability
density instead of a wavefunction. The density matrix is also given by a
gravitational path integral \cite{Page:1986vw}.
A no-boundary wavefunction can be a ``disconnected'' contribution
to the density matrix. In general, however, there will also be
``connected'' contributions which could arise, for instance, from
bra-ket wormholes \cite{Chen:2020tes}.
Recently, a conditional density matrix has been computed for
a complex geometry including bra-ket wormholes, and it was found
that the connected contribution dominates over the disconnected
contribution \cite{Fumagalli:2024msi}.

Contrary to the no-boundary wavefunction, trumpet wavefunctions
depend on a continuous parameter that labels the ``initial size'' of the
universe, with subsequent de Sitter expansion. In a recent paper, we
proposed a density matrix defined as sum over products of transition
amplitudes between an ``initial'' circle of size $\sqrt{h_0}$ and a ``final''
  circle of size $\sqrt{h}$ \cite{Buchmuller:2026qgp}. In the
  semiclassical regime this density matrix is consistent with the result obtained in 
\cite{Fumagalli:2024msi}. Remarkably, it leads to a probability
distribution that is flat in the universe size. In this paper, we
pursue this proposal further and compare in detail probability densities
obtained from the density matrix with those derived from the KG current.

The paper is organised as follows.
In section~\ref{sec:dm} we briefly review definition and properties of density matrices
defined by a path integral, following \cite{Page:1986vw}. In JT
gravity, the connected contribution to the density matrix is a
sum of double-trumpet amplitudes. The dependence of the trumpet
wavefunction on the path integral measure is discussed in section~\ref{sec:twf}, 
and the wavefunction consistent with holography is identified. 
Given trumpet and no-boundary wavefunctions, the corresponding
probability densities are compared in section~\ref{sec:pm}, and it is
shown that the density matrix singled out by holography leads to a
probability density flat in the universe size, as preferred by inflation.
It is well known that in dS JT gravity exact solutions of the WDW
equation have a singularity at a finite size of the universe 
\cite{Iliesiu:2020zld,Fanaras:2021awm,Buchmuller:2024ksd}. 
As discussed in section~\ref{sec:hs}, this singularity poses a serious
problem for a consistent definition of a Hilbert space for dS JT gravity.
For completeness, the connection between the KG inner product and
the group-averaging inner product is reviewed in appendix~\ref{A:ga},
following \cite{Held:2024rmg,Blommaert:2025bgd,Griguolo:2025kpi}.
We conclude in section~\ref{sec:conclusion}.

\section{Density matrix of de Sitter JT gravity}
\label{sec:dm}

Jackiw-Teitelboim gravity \cite{Teitelboim:1983ux,Jackiw:1984je}
in de Sitter space is defined by the Lorentzian action
\begin{equation}\label{SJT}
  iS_\text{dS}[g,\phi] = \frac{i}{4\pi} \int_{\mathcal{M}}d^2x \sqrt{g} \phi (R - 2\lambda^2) +
  \frac{i}{2\pi}\int_{\partial\mathcal{M}} d\theta \sqrt{h} \phi K \ .
\end{equation}
Here $g$, $R$, $\lambda^2$, $h$ and $K$ denote metric tensor, 
Ricci scalar and cosmological constant, and induced metric and extrinsic curvature on the boundary $\partial\mathcal{M}$, respectively. 
Compared to pure gravity with cosmological constant,
the action depends linearly on a dilaton field $\phi$.
In minisuperspace, which in 2d is just a gauge choice
\cite{Louis-Martinez:1993bge}, the metric with lapse function $N$, and
boundary metric $h$,
\begin{equation}\label{mLd2}
  ds^2 = -\frac{N^2}{h}dt^2 + h d\theta^2 \ , \quad 0\leq \theta < 2\pi \ ,
\end{equation}
yields the Lorentzian action 
\begin{equation}\label{ILjt}
  iI_\text{dS}[h,\phi] = i\int dt \left(-\frac{1}{2N} \dot{h} \dot{\phi} - N\lambda^2 \phi
    \right) \ .
  \end{equation}
This implies the Hamiltonian constraint 
    \begin{equation}\label{jtHC}
\left. \frac{\delta I_{\text{dS}}}{\delta N}\right|_{N=1} =
      \frac{1}{2}\dot{h} \dot{\phi} - \lambda^2 \phi = 0 
\end{equation}
on the phase-space coordinates $h$ and $\phi$, and the momenta
$\pi_\phi= -\dot{h}/2$ and $\pi_h=-\dot{\phi}/2$.

Quantization leads to the Wheeler-DeWitt (WDW) equation which, up to
factor ordering, reads
\begin{equation}\label{wdw0}
\left(2\partial_h\partial_\phi + \lambda^2 \phi\right)\psi(h,\phi)
\equiv \hat{\mathcal{H}}_{\text{WDW}}\psi(h,\phi) = 0 \ .
\end{equation}
Predictions of the theory depend on the choice of factor
ordering, which is related to the choice of measure
in the path integral \cite{Halliwell:1988wc}.

In general, the quantum system can be described by a density matrix
that is given by a path integral \cite{Page:1986vw}
\begin{equation}
  \rho(h,\phi;h',\phi') = 
 \int_{(h',\phi')}^{(h,\phi)}[Dg][D\phi']\exp{(iS_G[g,\phi'])} \ ,
\end{equation}
where $(h,\phi)$ and $(h',\phi')$ denote metric and dilaton field on
the two boundary surfaces. The density satisfies a WDW equation,
\begin{equation}
 \mathcal{H}_{\text{WDW}} \rho(h,\phi;h',\phi') = \mathcal{H}'_{\text{WDW}}
 \rho(h,\phi;h',\phi') = 0 \ ,
\end{equation}
and can therefore be written as
\cite{Page:1986vw}
\begin{equation}\label{rhomix}
  \rho(h,\phi;h',\phi') = \sum_{i,j} C_{i,j} \psi_i(h,\phi)
  \psi^*_i(h',\phi') \ , \quad C_{i,j} = C^*_{j,i} \ ,
\end{equation} 
with wavefunctions $\psi_i$ satisfying the WDW equation
$\mathcal{H}\psi_i = 0$. Generically, this density matrix describes a
mixed state.
In case of a pure state only a
single wavefunction contributes to the sum. This is then a
``wavefunction of the universe'', like the no-boundary wavefunction
\cite{Hartle:1983ai}.

\begin{figure}[t]
 \centering
 \includegraphics[width = 0.8 \textwidth]{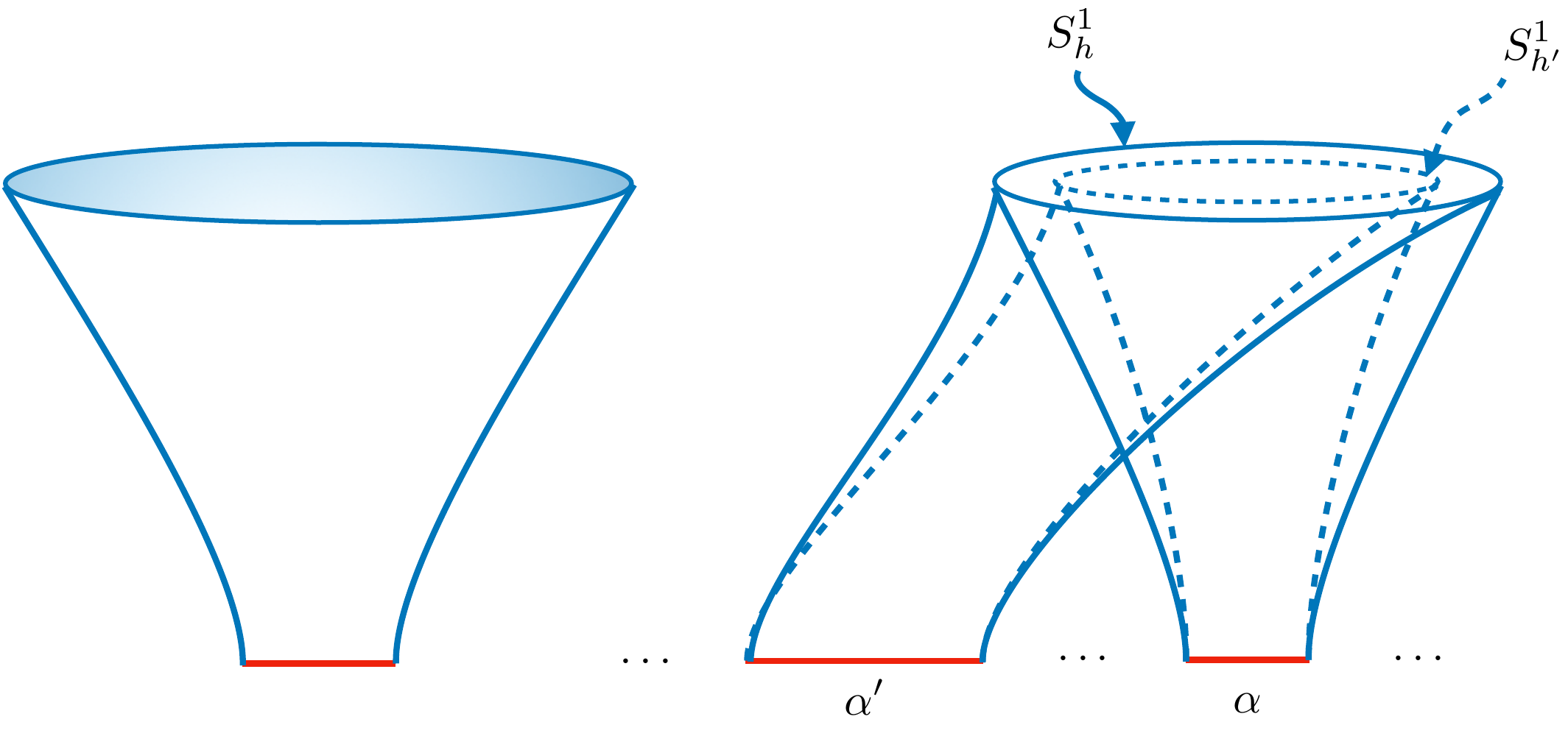} 
% \hspace{2cm}
% \includegraphics[width = 0.44 \textwidth]{DT} 
 \caption{Left: single Lorentzian trumpet. Right: two Lorentzian
     double-trumpets for two different values of $\alpha=\sqrt{h_0}$.}
\label{fig:WF1}
\end{figure}

In the ``half reduction'' version \cite{Svesko:2022txo} of dS JT gravity, field space is
bounded by the two half axes $\phi = 0, h > 0$ and $h = 0, \phi > 0$.
One has connected geometries contributing to the density matrix
and possibly also a disconnected geometry, depending on how the path
integral is defined (see fig.~\ref{fig:WF1}).
The connected contribution has the form of a double-trumpet where each
trumpet is a section of a hyperboloid connecting a geodesic circle of
size $\alpha=\sqrt{h_0}$ at $\phi = 0$ with a circle of size $\sqrt{h}$ at
$\phi_b$. Two trumpets are glued together to a double-trumpet at the
boundary $\phi = 0$.
The contribution to the density matrix then has the form
\eqref{rhomix} with trumpet wavefunctions $\psi_i$. 

\section{Trumpet wavefunctions}
\label{sec:twf}

The JT gravity action \eqref{SJT} is linear in the dilaton field
$\phi$. Hence, $\phi$ can be integrated out, which enforces the
constraint $R=2\lambda^2$ on metrics that contribute to the
gravitational path integral. Geometrically, these correspond to pieces
of a de Sitter hyperboloid. Consider now a ``trumpet configuration''
with two boundaries one of which is a geodesic surface that has zero extrinsic
curvature (see fig.~\ref{fig:WF1}). More complicated metrics can be obtained by glueing together
different trumpets along geodesic surfaces. 

This program has first been carried out in \cite{Saad:2019lba} for
Euclidean $\text{AdS}_2$.  With $t = -i\rho$, the metric \eqref{mLd2} reads
\begin{equation}
  ds^2 = \frac{N^2}{h}d\rho^2 + h d\theta^2 \ , \quad 0\leq \theta < 2\pi \ .
\end{equation}
Here $h$ and $\phi$ are functions of the ``radial'' coordinate $\rho$, and
$\theta$ plays the role of a  Euclidean time coordinate in the boundary
CFT. Substituting in addition $\lambda$ by $i\lambda$, the Lorentzian action
$iI_{\text{dS}}$ becomes the Euclidean $\text{AdS}_2$ action
\begin{equation}
  I_\text{EAdS}[h,\phi] = \int d\rho \left(\frac{1}{2N} h' \phi' + N\lambda^2 \phi
    \right) \ .
  \end{equation}
For this theory the trumpet partition function has been calculated in
the limit $h,\phi \rightarrow \infty$, $\sqrt{h}/\phi$ fixed, by
integrating over the Schwarzian boundary fluctuations. This limit
requires the addition of a counterterm which amounts to the
replacement $K \rightarrow K-1$ in the action \eqref{SJT}.
In case of a geodesic boundary of length $2\pi \alpha$ one finds \cite{Saad:2019lba},
\begin{equation}\label{ZTE}
 Z_T\left(\frac{\phi}{\sqrt{h}},\alpha\right) = \left(\frac{\phi}{2\lambda\sqrt{h}}\right)^{1/2} 
\exp{\left(-\frac{\lambda\phi \alpha^2}{2\sqrt{h}}\right)} \ .
\end{equation}

An analogous calculation has been performed for Lorentzian $\text{dS}_2$
\cite{Cotler:2019nbi}. The corresponding ``future'' and ``past''
trumpet amplitudes can be obtained from the Euclidean partition
function \eqref{ZTE} by analytic continuation. Replacing $\lambda$ by $-i\lambda$
yields \cite{Cotler:2019nbi}
\begin{equation}\label{ZT+}
 Z_{T+}\left(\frac{\phi}{\sqrt{h}},\alpha\right) =
 \left(\frac{\phi}{-2i\lambda\sqrt{h}}\right)^{1/2}
 \exp{\left(\frac{i\lambda\phi \alpha^2}{2\sqrt{h}}\right)} \ .
\end{equation}
The substitution $\lambda \rightarrow i\lambda$ in \eqref{ZTE} leads
  to $Z_{T-} = Z_{T+}^*$. 
Without the counterterm in the action \eqref{SJT} one finds for the trumpet
wavefunction at large, finite $\sqrt{h}$, $\phi$ (see \cite{Buchmuller:2024ksd}),
\begin{equation}\label{T+sc}
 \psi_{T+}(h,\phi;\alpha) \sim \left(\frac{\phi}{\sqrt{h}}\right)^{1/2}
 \exp{\left(-i\lambda\phi\left(\sqrt{h}-\frac{\alpha^2}{2\sqrt{h}}\right)\right)} \ ,
\end{equation}
which satisfies the WDW equation with Henneaux factor
ordering\footnote{See
    \cite{Henneaux:1985nw,Louis-Martinez:1993bge,Iliesiu:2020zld}, and
    \cite{Franken:2025unn} for a discussion based on the Hartle-Hawking
    inner product.},
\begin{equation}\label{wdwH} 
  \left(2h^{1/2}\partial_hh^{-1/2}\partial_\phi
    + \lambda^2 \phi\right)\psi_{T+} \equiv
\mathcal{H}_{\text{WDW}}\psi_{T+} =  0 \ ,
\end{equation}
up to terms $\mathcal{O}(1/\chi^3)$, with $\chi = \sqrt{h},\phi$. 

The trumpet partition function for finite $\sqrt{h}$, $\phi$ has first
been computed in \cite{Iliesiu:2020zld} by summing an infinite series
of extrinsic curvature terms. The obtained result reads
\begin{equation}\label{ZTfinite}
  Z_T(h,\phi;\alpha) = \frac{\sqrt{h}\phi}{\sqrt{h-\alpha^2}} e^{-\lambda\phi\sqrt{h}}
  K_1(-\lambda\phi\sqrt{h-\alpha^2}) \ .
\end{equation}
Removing the counterterm factor $\exp{(-\lambda\phi\sqrt{h})}$, and
analytically continuing
to $\text{dS}_2$, one finds the trumpet wavefunction\footnote{Here we have
  dropped a factor $-\pi$ to account for different conventions.}
\begin{equation}\label{T+finite}
  \psi_{T+}(h,\phi;\alpha) = \frac{\sqrt{h}\phi}{2\sqrt{h-\alpha^2}}
  H^{(2)}_1(\lambda \phi(h-\alpha^2)^{1/2})  \ , \quad h > \alpha^2 \ .
\end{equation}
$\psi_{T+}$ is an exact solution of the WDW equation \eqref{wdwH}.
It is also an eigenfunction of the operator
\begin{equation}
  \mathcal{G} = \partial_\phi^2 - \phi^{-1}\partial_\phi + \lambda^2 h
\end{equation}
with eigenvalue $\lambda^2\alpha^2$,
\begin{equation}
\mathcal{G} \psi_{T+}(h,\phi;\alpha) = \lambda^2\alpha^2
\psi_{T+}(h,\phi;\alpha) \ .
\end{equation}
$\mathcal{G}$ and the Hamiltonian \eqref{wdwH} have the commutator
\begin{equation}
  [ \mathcal{H}_{\text{WDW}},\mathcal{G}]
  = \phi^{-2} \mathcal{H}_{\text{WDW}}  \ .
  \end{equation}
  Hence, $\mathcal{G}$ and the Hamiltonian commute on the space
  of solutions of the WDW equation, which means that $\mathcal{G}$
  is an observable. The eigenvalue $\alpha$ corresponds to the size
  of the geodesic circle.

Note that both $Z_T$ and $\psi_{T+}$ have a pole at $h=\alpha^2$, where
$\psi_{T+} \sim -i\sqrt{h}/(h-\alpha^2)$.
Summing the trumpet wavefunctions of outgoing and incoming branches,
$\psi_T = \psi_{T+} + \psi_{T-}$,
one obtains the real wavefunction
\begin{equation}\label{Tfinite}
  \psi_{T}(h,\phi;\alpha) = \frac{\sqrt{h}\phi}{\sqrt{h-\alpha^2}}
 J_1(\lambda \phi(h-\alpha^2)^{1/2})  \ ,
\end{equation}
which is regular at $h=\alpha^2$ and vanishes at $\phi = 0$.

So far we have considered trumpet partition functions obtained from
the boundary Schwarzian path integral. Alternatively, one can compute
the bulk path integral with Dirichlet boundary conditions $(h,\phi)$
on one surface and the boundary condition $(\alpha^2,\pi_\phi=0)$
on the geodesic surface, using the fact
that there the extrinsic curvature $\sim \pi_\phi$
vanishes. This has recently been worked out  for $\text{EAdS}_2$ in
\cite{Griguolo:2025kpi}. The result, analytically continued to
$\text{dS}_2$, reads
\begin{equation}\label{tildeT}
\tilde{\psi}_T(h,\phi;\alpha) =
\frac{1}{\lambda\sqrt{h-\alpha^2}}J_1(\lambda\phi\sqrt{h-\alpha^2}) \ .
\end{equation}
This wavefunction satisfies a WDW equation with non-flat ordering of
the form \cite{Griguolo:2025kpi},
\begin{equation}\label{wdwk}
  \left(2\partial_h\phi^{-1}\partial_\phi\phi
    +\lambda^2 \phi\right)\tilde{\psi}_{T}
 \equiv \tilde{\mathcal{H}}_{\text{WDW}}\tilde{\psi}_{T}
  = 0 \ ,
\end{equation}
and it is an eigenfunction of the operator
\begin{equation}
\tilde{\mathcal{G}} = \partial^2_\phi + \phi^{-1} \partial_\phi -
\phi^{-2} + \lambda^2 h
\end{equation}
with eigenvalue $\lambda^2\alpha^2$. As in the case of Henneaux
factor ordering, $\tilde{\mathcal{G}}$ commutes with the Hamiltonian
on-shell, i.e., $[\tilde{\mathcal{H}}_{\text{WDW}}, \tilde{\mathcal{G}}]
=\phi^{-2}\tilde{\mathcal{H}}_{\text{WDW}}$.

The wavefunction $\tilde{\psi}_T$ differs from $\psi_T$ by a factor
$\lambda\sqrt{h}\phi$ which can be accounted for by
a modified path integral measure \cite{Halliwell:1988wc,Griguolo:2025kpi}.
We conclude that the trumpet wavefunctions \eqref{T+finite}
and \eqref{Tfinite}, computed
from the boundary Schwarzian path integral, agree with the wavefunctions
computed from a bulk path integral with appropriate measure. By
construction, these trumpet wavefunctions are consistent with holography.

A further interesting trumpet wavefunction has
been introduced in \cite{Held:2024rmg} and further studied in
\cite{Blommaert:2025bgd,Griguolo:2025kpi},
 \begin{equation}\label{tHM}
   \hat{\psi}_{T+}(h,\phi;\alpha) 
  = \frac{1}{\lambda\sqrt{h - \alpha^2}}e^{-i(\lambda\phi\sqrt{h-\alpha^2})} \ .
\end{equation}
$\hat{\psi}_{T+}(h,\phi;\alpha)$ is a solution to the WDW equation
\eqref{wdw0} with flat ordering.
It is an eigenfunction of the operator $\hat{\mathcal{G}} =
  \partial_\phi^2 + \lambda^2h$,
which commutes with the Hamiltonian, with eigenvalue
$\lambda^2\alpha^2$ \cite{Held:2024rmg},
\begin{equation}
  \hat{\mathcal{G}} \hat{\psi}_{T+}(h,\alpha) = \left(
    \partial_\phi^2 + \lambda^2h\right)\hat{\psi}_{T+}(h,\phi;\alpha)
  = \lambda^2\alpha^2 \hat{\psi}_{T+}(h,\phi;\alpha) \ .
\end{equation}
The wavefunction \eqref{tHM} can also be obtained from the path
integral by modifying the treatment of the zero mode of the dilaton
field \cite{Griguolo:2025kpi}.

The trumpet wavefunctions $\psi_T$, $\tilde{\psi}_T$ and
  $\hat{\psi}_T$ are all transition amplitudes between a geodesic
  circle and a second circle with Dirichlet boundary conditions.
  They reflect different definitions of the bulk path
  integral. However, only one of them, $\psi_T$, has holographic
  asymptotic behaviour, consistent with Schwarzian boundary
  fluctuations. Concerning the bulk path integral, this singles
  out Henneaux factor ordering.

The real Hartle-Hawking no-boundary wavefunction with Schwarzian
asymptotic behaviour is obtained by
convolving the trumpet amplitude with a ``cap amplitude'' that is
singular at the critical size $\alpha = \sqrt{h_c} \equiv \lambda^{-1}$
\cite{Buchmuller:2024ksd,Griguolo:2025kpi},
\begin{equation}\label{psirho}
  \Psi_{\text{NB}}(h,\phi) = \int dh' \psi_T(h,\phi;\sqrt{h'})\rho_{h_c}(h') \ ,
  \quad \rho_{h_c}(h') = -\partial_{h'}\delta(h'-h_c) \ . 
\end{equation}
Writing the trumpet wavefunction \eqref{Tfinite} in the compact form\footnote{In the
  classification \cite{Buchmuller:2024ksd} of solutions of the WDW equation 
  w.r.t. factor ordering and singularity at the geodesic boundary,
  $\psi_T(h,\phi;\alpha^2) = \Psi^{(1,1)}(h,\phi;\alpha^2)$.}
\begin{equation}\label{Treal}
  \psi_T(h,\phi;\alpha) = -2\lambda^{-1}\sqrt{h}
  \partial_h  J_0(\lambda \phi(h-\alpha^2)^{1/2}) \ , 
\end{equation}
one obtains for the no-boundary wavefunction,\footnote{The
  no-boundary wavefunction
  corresponds to $\Psi_{\text{NB}}(h,\phi) = \Psi^{(1,2)}(h,\phi;h_c)$
  in \cite{Buchmuller:2024ksd}.}
  \begin{equation}
 \Psi_{\text{NB}}(h,\phi)
  = -2\lambda^{-1}\sqrt{h}\partial_h^2 J_0(\lambda\phi(h-h_c)^{1/2}) \ .
\end{equation}
At large $h$ and $\phi$, the Schwarzian scaling is obtained,
\begin{equation}\label{HHwf}
\Psi_{\text{NB}}(h,\phi) \sim 
   \left(\frac{\phi}{\sqrt{h}}\right)^{3/2}
   \cos{\left(\lambda\phi\sqrt{h}\left(1 - \frac{h_c}{2h}\right)\right)}\ .
 \end{equation}

For the trumpet wavefunction $\hat{\psi}_{T} = \hat{\psi}_{T+} +
\hat{\psi}^*_{T+}$, the convolution \eqref{psirho}
yields the no-boundary wavefunction
 \begin{equation}\label{NBHM}
  \hat{\Psi}_{\text{NB}}(h,\phi;h_c) = \partial_h \hat{\psi}_T(h,\phi;h_c)
  \ , \quad h > h_c \ .
\end{equation}
At large $h$ and $\phi$, this no-boundary wavefunction becomes
\begin{equation}\label{HMsc}
\hat{\Psi}_{\text{NB}}(h,\phi) \sim 
   \left(\frac{\phi}{h}\right)
   \cos{\left(\lambda\phi\sqrt{h}\left(1 - \frac{h_c}{2h}\right)\right)}\ .
 \end{equation}
We note that for these states the prefactor no longer takes the form of a monomial in $\phi/\sqrt{h}$.

\section{Probability measures}
\label{sec:pm}

In quantum cosmology one is interested in computing a probability
distribution for the size of the universe. Given a classical action of
gravity and matter, and a procedure of quantization, how does one
identify a ground state, a ``state of minimal excitation'', and
determine the corresponding probability distribution?

In nonrelativistic quantum mechanics the diagonal element of a
density matrix plays the role of a probability density. As discussed
in section~\ref{sec:dm}, in quantum gravity
one has to consider a conditional density matrix. Imposing in JT
gravity a condition on the dilaton field, $\phi=\phi'=\phi_b$, one obtains
from eq.~\eqref{rhomix}
\begin{equation}
  \rho(h,\phi_b;h',\phi_b) = \sum_{i,j} C_{i,j} \psi_i(h,\phi_b)
  \psi^*_i(h',\phi_b) \ , \quad C_{i,j} = C^*_{j,i} \ ,
\end{equation} 
with wavefunctions $\psi_i$ satisfying a WDW equation
$\mathcal{H}_{\text{WDW}}\psi_i = 0$. As we argued in the previous section,
a surface with two boundaries, $(h,\phi_b)$ and $(h',\phi_b)$,
is obtained by glueing the geodesic surfaces of two trumpets together
(see \cite{Saad:2019lba,Cotler:2019nbi}). For the density matrix
of outgoing universes this yields the expression
\begin{equation}\label{cdm}
  \rho_+(h,\phi_b;h',\phi_b) = \int_{0}^{\kappa}\alpha d\alpha \psi_{T+}(h,\phi_b;\alpha)
  \psi_{T+}(h',\phi_b;\alpha)^* \ , \quad h,h' > \kappa^2,
\end{equation} 
where we have integrated the size of the geodesic surfaces over a
compact domain $[0,\kappa]$. In order to avoid the singularities at $h
= \alpha^2$, $h' = \alpha^2$, we require $h,h' > \kappa^2$.

Using eq.~\eqref{T+sc}, the density matrix 
at large $h,h'\gg \kappa$ becomes
\begin{equation}\label{rho+}
 \rho_+(h,\phi_b;h',\phi_b) = \int_{0}^{\kappa}\alpha d\alpha
 \rho^{(\alpha,\phi_b)}_+(h,h') \ , 
\end{equation}
with
\begin{equation}\label{rho+sc}
\begin{split}
  \rho^{(\alpha,\phi_b)}_+(h,h') &\sim 
   \left(\frac{\phi_b^2}{\sqrt{h}\sqrt{h'}}\right)^{1/2} \\
&\quad\times\exp\left(-i\lambda\phi_b\left(\sqrt{h} - \sqrt{h'} -
 \frac{\alpha^2}{2}\left(\frac{1}{\sqrt{h}}-\frac{1}{\sqrt{h'}}\right)\right)\right)  \ .
\end{split}
     \end{equation} 
This is precisely the conditional density matrix obtained in
\cite{Fumagalli:2024msi},
based on a complex geometry with bra-ket wormholes
\cite{Chen:2020tes}. It is the finite-cutoff version of the
double-trumpet amplitude at infinite cutoff which has been
computed in \cite{Saad:2019lba,Cotler:2019nbi}.
The density matrix \eqref{rho+sc} differs by a factor $\phi_b^2$ from
the one proposed in \cite{Buchmuller:2026qgp}. This does not
change the probability distribution in $h$ at fixed $\phi_b$. However,
it does affect
the interpretation of the density matrix, as we shall discuss below.

In addition to the connected contribution to the density matrix there
can be a disconnected contribution given by the Hartle-Hawking
wavefunction,
\begin{equation}\label{rho+dis}
 \rho^{\text{dis}}_+(h,\phi_b;h',\phi_b) = \Psi_{\text{NB}+}(h,\phi_b)
 \Psi_{\text{NB}+}(h',\phi_b)^* \ .
 \end{equation}
From eq.~\eqref{HHwf} one obtains for $h,h' \gg h_c$,
 \begin{equation}
   \begin{split}
   \rho^{\text{dis}}_+(h,\phi_b;h',\phi_b)
   &\sim
   \left(\frac{\phi_b^2}{\sqrt{h}\sqrt{h'}}\right)^{3/2} \\
&\quad\times\exp\left(-i\lambda\phi_b\left(\sqrt{h} - \sqrt{h'} -
 \frac{h_c}{2}\left(\frac{1}{\sqrt{h}}-\frac{1}{\sqrt{h'}}\right)\right)\right)  \ .
\end{split}
     \end{equation}   
Compared to the connected contribution \eqref{rho+sc}, $\alpha^2$ is now
fixed to $h_c$, and the falloff with size $\sim (hh')^{-1/4}$ becomes steeper
$\sim (hh')^{-3/4}$.

The diagonal element of the density matrix \eqref{cdm},
\begin{equation}\label{cp}
  p_+(h|\phi_b) \equiv \rho_+(h,\phi_b;h,\phi_b)
  = \int_{0}^{\kappa}\alpha d\alpha \psi_{T+}(h,\phi_b;\alpha)
  \psi_{T+}(h,\phi_b;\alpha)^* \ , 
\end{equation} 
is usually interpreted as the probability density to find a
universe of size $\sqrt{h}$ in an ensemble of universes under the
condition that the dilaton takes the value $\phi=\phi_b$. For
arbitrary $\sqrt{h} > \kappa$ this interpretation is questionable.
In the semiclassical regime, however, for $\sqrt{h} \gg \kappa$,
the interpretation is widely accepted. $p_+(h|\phi)$ then
corresponds to a probability measure on the space of classical
trajectories of the system (see, for example, \cite{Vilenkin:1993ym,Halliwell:1992cj}).

\begin{figure}[t!]
 \centering
 \includegraphics[width = 0.7 \textwidth]{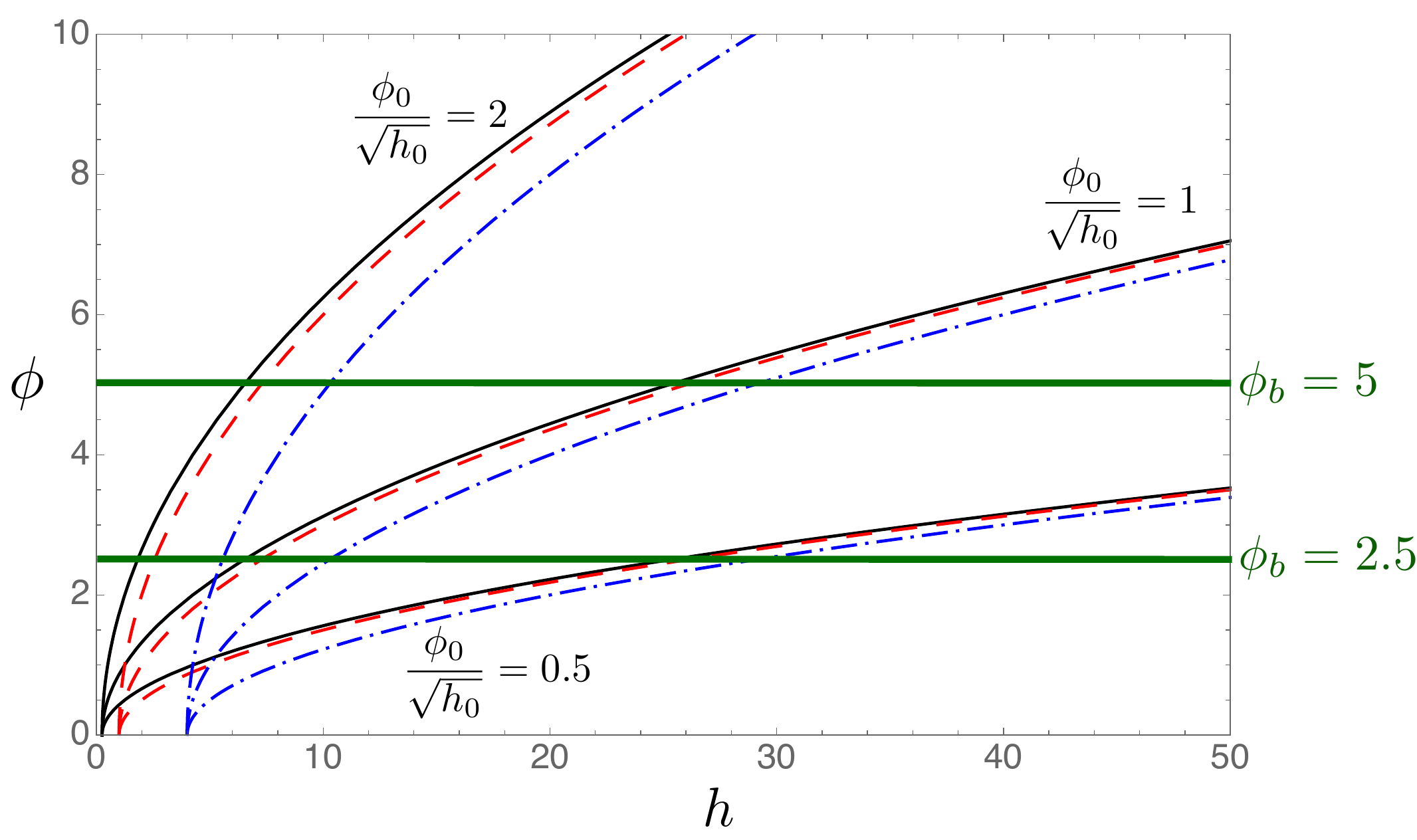} 
 \caption{Congruences of classical trajectories pertaining to fixed $\phi_0/\sqrt{h_0}$.}
\label{fig:trajects}
\end{figure}

In de Sitter JT gravity one is interested in the
question how late time classical trajectories emerge from an initial
quantum state. Solutions of the classical equations of motion are
characterized by some integration constants. In the semiclassical
regime a probability density computed from a quantum state should
be constant on a given classical trajectory but it should take
different values for different trajectories. It can then be viewed
as a probability density on the space of integration
constants which parametrize the possible classical trajectories
of the system \cite{Halliwell:1992cj}.

Consider now dS JT gravity from half reduction, with field values
$h>0$, $\phi > 0$. The trumpet wavefunction is the transition
amplitude between a geodesic circle of size $\sqrt{h_0}$ at $\phi=0$
and a second circle of size $\sqrt{h}$ at $\phi_b$.
Classical trajectories $(h,\phi)$ are characterized by two
constants, $h_0$ and $\phi_0$  (see, for example, \cite{Buchmuller:2026qgp}), 
\begin{equation}\label{hphi}
  \phi = \phi_0 \Dh h_0^{-1/2} \ , \quad \text{with} \quad
 \Dh = (h-h_0^2)^{1/2} \ .
\end{equation}
A trajectory $(h_0,\phi_0)$ intersects the boundary $\phi=\phi_b$
at $h_b$. Changing the boundary, i.e. $\phi_b$, leads to a different
value of $h_b$.
For $h_b \gg h_0$, one has on the trajectory $(h_0,\phi_0)$,
\begin{equation}
\frac{\phi_b}{\sqrt{h_b}} \simeq \frac{\phi_0}{\sqrt{h_0}} \ .
\end{equation}
From \eqref{rho+sc} one reads off that 
the probability density $p_+$ on the trajectory $(h_0,\phi_0)$ is indeed
constant and determined by the initial conditions,
  \begin{equation}\label{p+con}
    p_+(h_b|\phi_b) \sim  \frac{\phi_b}{\sqrt{h_b}}  \simeq
    \frac{\phi_0}{\sqrt{h_0}} \ .
    \end{equation}
Different trajectories, with different values of $\phi_0/\sqrt{h_0}$,
intersect the boundary $\phi=\phi_b$ at different values of $h_b$.
For a visualization see figure~\ref{fig:trajects}.
In this way the distribution in $h$ for fixed $\phi_b$ becomes a
distribution on the congruence of classical trajectories. 

The situation is different for the
  density matrix $\tilde{\rho}_+ \sim \phi_b^{-2}\rho_+$
    defined in \cite{Buchmuller:2026qgp}. There, for a given trajectory
    $(h_0,\phi_0)$ the probability density reads
      \begin{equation}
        \tilde{p}_+(h_b|\phi_b)  = \tilde{\rho}_+(h_b,\phi_b;h_b,\phi_b)
        \simeq  \phi_b^{-2}\frac{\phi_0}{\sqrt{h_0}}\  .
    \end{equation}
    This expression is not constant on the trajectory $(h_0,\phi_0)$
    but varies with $\phi_b$. Hence, the
    density matrix defined in \cite{Buchmuller:2026qgp} does not yield
    a probability measure on classical trajectories.
    
 Like the connected contribution to the density matrix
    also the disconnected contribution \eqref{rho+dis} leads to a
    probability density that is constant on a single trajectory since
    the Hartle-Hawking wavefunction only depends on the ratio
    $\phi/\sqrt{h}$,
        \begin{equation}\label{pdis1}
      p_+^{\text{dis}}(h_b|\phi_b) \propto 
      \Psi_{\text{NB}+}(h_b,\phi_b)\Psi_{\text{NB}+}(h_b,\phi_b)^*
      \sim \left(\frac{\phi_b}{\sqrt{h_b}}\right)^3
      \simeq \left(\frac{\phi_0}{\sqrt{h_0}}\right)^3\ .
      \end{equation}
    Similarly, a probability density defined by a
    Klein-Gordon current density of a no-boundary wavefunction is
    constant on a single trajectory, 
    \begin{equation}\label{pdis2}
      p^{\text{KG}}_{+}(h_b|\phi_b) \propto \text{Im}
      \left(\Psi_{\text{NB}+}(h,\phi_b)\partial_h
        \Psi_{\text{NB}+}(h,\phi_b)^*\right)|_{h=h_b}
      \sim \left(\frac{\phi_b}{\sqrt{h_b}}\right)^4
      \simeq \left(\frac{\phi_0}{\sqrt{h_0}}\right)^4\ .  
    \end{equation}

    The corresponding probability densities for the trumpet
    wavefunction $\hat{\psi}_{T+}$ can be read off from eqs.~\eqref{tHM}
    and \eqref{HMsc},
    \begin{align}
    \hat{p}_+(h_b|\phi_b)  &= \hat{\rho}_+(h_b,\phi_b;h_b,\phi_b)
      \sim h_b^{-1}  \simeq
      \phi_b^{-2}\left(\frac{\phi_0}{\sqrt{h_0}}\right)^2\ , \\
\hat{p}_+^{\text{dis}}(h_b|\phi_b) &\propto 
      \hat{\Psi}_{\text{NB}+}(h_b,\phi_b)\hat{\Psi}_{\text{NB}+}(h_b,\phi_b)^*
      \sim \left(\frac{\phi_b}{h_b}\right)^2 \simeq  
\phi_b^{-2}\left(\frac{\phi_0}{\sqrt{h_0}}\right)^4\ , \\
\hat{p}^{\text{KG}}_{+}(h_b|\phi_b) &\propto \text{Im}
      (\hat{\Psi}_{\text{NB}+}(h,\phi_b)\partial_h
        \hat{\Psi}_{\text{NB}+}(h,\phi_b)^*)|_{h=h_b}
      \sim \frac{\phi_b^3}{\sqrt{h_b}^5}
      \simeq \phi_b^{-2}\left(\frac{\phi_0}{\sqrt{h_0}}\right)^5\ .  
\end{align}
Clearly, none of these probability densities defines a probability
measure on trajectories.

The conditional probability densities \eqref{pdis1} and \eqref{pdis2},
\begin{equation}
p_+^{\text{dis}}(h|\phi_b) \sim
\left(\frac{\phi_b}{\sqrt{h}}\right)^3\ , \quad
    p^{\text{KG}}_{+}(h|\phi_b) 
      \sim \left(\frac{\phi_b}{\sqrt{h}}\right)^4 \ ,
    \end{equation}
 are large-$h$ normalizable. However, this has the well known
 consequence that large universes are disfavoured
 \cite{Hartle:2008ng,Maldacena:2024uhs}, which is inconsistent with
 inflation. The only probability density that defines a measure on classical
 trajectories and is consistent with inflation, is given by the
 connected contribution to the density matrix \eqref{p+con},
   \begin{equation}
    p_+(h|\phi_b) \sim  \frac{\phi_b}{\sqrt{h}}  \ .
    \end{equation}
With $dh = 2\sqrt{h}d\sqrt{h}$, this yields a flat probability distribution in the
size of the universe $a=\sqrt{h}$, $dP_+ = p_+dh \sim \phi_b d\sqrt{h}$.
The probability density $p_+$ is unnormalizable, so that one can
only compute relative probabilities.

\section{Hilbert space of de Sitter JT gravity}
\label{sec:hs}

The wavefunctions discussed in section~\ref{sec:twf} are all solutions
of a WDW equation and should therefore represent physical states which
are elements of a Hilbert space whose structure may depend on the
chosen operator ordering. A Hilbert space requires a positive-definite
inner product. In the following we compare the
Klein-Gordon (KG) inner product for the different wavefunctions considered
in section~\ref{sec:twf}. The connection to the group averaging
inner product is reviewed in appendix~\ref{A:ga}.

Solutions of a WDW equation with kinetic term $\Delta =
\sqrt{f}^{-1}\partial_\alpha\left(\sqrt{f}f^{\alpha\beta}\partial_\beta\right)$,
with $\alpha,\beta = h, \phi$, have a conserved KG current
\begin{equation}
  j^\alpha = \frac{i}{2}f^{\alpha\beta}\left(\psi^*\partial_\beta\psi
    - \psi\partial_\beta\psi^*\right) \ , \quad \partial_\phi j_h +
  \partial_h j_\phi = 0 \ .
\end{equation}  
The KG inner product is obtained by integrating the Hodge-dual over a
line. Choosing $\Sigma = \{\phi=\phi_b, h > 0\}$, one obtains for an
outgoing wavefunction $\psi_+$,
\begin{equation}\label{ip1}
\langle \psi_+, \psi_+ \rangle = \int_\Sigma \star j = \int dh j^+_h \ .
\end{equation}
In the case of `half reduction' the geodesic surface is located at
$\phi = 0$. Its size is $\alpha \equiv \sqrt{h_0}$ for trumpet
wavefunctions and $\alpha = \sqrt{h_c} = \lambda^{-1}$ for the
no-boundary wavefunction. Using the correspondence~\cite{Buchmuller:2024ksd}
$\Psi_{\text{NB}+}(h,\phi_b;h_c^{1/2})
= \Psi^{(1,2)}(h,\phi_b;h_c)$ and $\psi_{T+}(h,\phi_b;\alpha)
= \Psi^{(1,1)}(h,\phi_b;\alpha^2)$, one finds
for the current density $j^+_h$ at large $h$, as well as close to the
singularity,
\begin{align}
  \Psi_{\text{NB}+}(h,\phi_b;h_c^{1/2}): \quad & j^+_h \sim \frac{h \phi_b^4}{(h-h_c)^3}\ ,
     \quad h \sim h_c \ ; \quad  j^+_h \sim \frac{\phi_b^4}{h^2}\ , \quad h \gg h_c \ , \\
  \psi_{T+}(h,\phi_b;\alpha):\quad & j^+_h \sim \frac{h \phi_b^2}{(h-\alpha^2)^2}\ ,
                                     \quad h \sim \alpha^2 \ ; \quad  j^+_h \sim \frac{\phi_b^2}{h}\ ,
                                     \quad h \gg \alpha^2 \ . 
\end{align}
Clearly, both wavefunctions are not normalizable because of the
singularity at $h = h_c$ and $h = \alpha^2$, respectively. Contrary to
$\Psi_{\text{NB}+}$, the trumpet wavefunction
$\psi_{T+}$ is also not large-$h$ normalizable. Hence,
$\Psi_{\text{NB}+}$ and
$\psi_{T+}$ do not fulfill the criterium of square-integrability of
Hilbert-space wavefunctions.

The trumpet amplitude introduced in \cite{Held:2024rmg},
 \begin{equation}\label{tHMr}
   \hat{\psi}_T(h,\phi;\alpha) 
  = \frac{1}{\pi\lambda\sqrt{h - \alpha^2}}\cos{(\lambda\phi\sqrt{h-\alpha^2})} \ ,
\end{equation}
is a solution to the WDW equation \eqref{wdw0}, with ``flat'' operator
  ordering leading to a flat metric in field space. Unlike the
wavefunctions considered in \cite{Buchmuller:2024ksd}, it does not
have Schwarzian behaviour at large $h$, $\phi$. Furthermore, it
diverges at $h \sim \alpha^2$, contrary to the real wavefunction
$\psi_T$ (see \eqref{Treal}). 
As recently shown in \cite{Griguolo:2025kpi}, in `full reduction' JT
gravity \cite{Svesko:2022txo} at infinite cutoff, where $\phi$ varies
from $-\infty$ to $+\infty$, $\hat{\psi}_T$ is given by a Fourier
transform of the transition amplitude $\langle
h,\phi|\alpha^2,0\rangle$ considered in \cite{Buchmuller:2026qgp}, 
\begin{equation}
  \hat{\psi}_T(h,\phi;\alpha) = \langle h,\phi|\alpha^2,0\rangle_{\text{FT}} \ ,
\end{equation}
with
\begin{equation}
  \langle h,\phi|\alpha^2,k'\rangle_{\text{FT}} =
\int_{-\infty}^{\infty}d\phi' e^{-ik'\phi'} \langle h,\phi|\alpha^2,\phi'\rangle \ .
\end{equation}
Hence, $\hat{\psi}_T$ is the transition amplitude from a surface of
size $\alpha$ and zero extrinsic curvature to a surface $(h,\phi)$,
whereas the amplitude 
\begin{equation}
\langle h,\phi|\alpha^2,0\rangle =
\frac{i}{2\pi}J_0(\lambda \phi(h-\alpha^2)^{1/2}) 
\end{equation}
corresponds to Dirichlet boundary conditions on both surfaces \cite{Halliwell:1990tu,Buchmuller:2024ksd}.
Convolving $\hat{\psi}_T(h,\phi;\alpha)$ with the cap amplitude, like in
eq.~\eqref{psirho}, one obtains the no-boundary wavefunction
\begin{equation}
  \hat{\Psi}_{\text{NB}}(h,\phi;h_c^{1/2}) = \partial_h \hat{\psi}_T(h,\phi;h_c^{1/2})
  \ , \quad h > h_c \ .
\end{equation}
From eqs.~\eqref{tHM} and \eqref{NBHM} one reads off the
expressions for $j^+_h$ at large $h$ and close to the
singularity, 
\begin{align}
  \hat{\Psi}_{\text{NB}+}(h,\phi_b;h_c^{1/2}): \quad & j^+_h \sim \frac{\phi_b}{(h-h_c)^2}\ ,
     \quad h \sim h_c \ ; \quad  j^+_h \sim \frac{\phi_b}{h^2}\ , \quad h \gg h_c \ , \\
  \hat{\psi}_{T+}(h,\phi_b;\alpha):\quad & j^+_h \sim \frac{\phi_b}{(h-\alpha^2)^{3/2}}\ ,
     \quad h \sim \alpha^2 \ ; \quad  j^+_h \sim \frac{\phi_b}{h^{3/2}}\ , \quad h \gg \alpha^2 \ . 
\end{align}
Again, both wavefunctions are not normalizable because of the
singularity at $h = h_c$ and $h = \alpha^2$, respectively.

\begin{figure}[t!]
 \centering
 \includegraphics[width = 0.35 \textwidth]{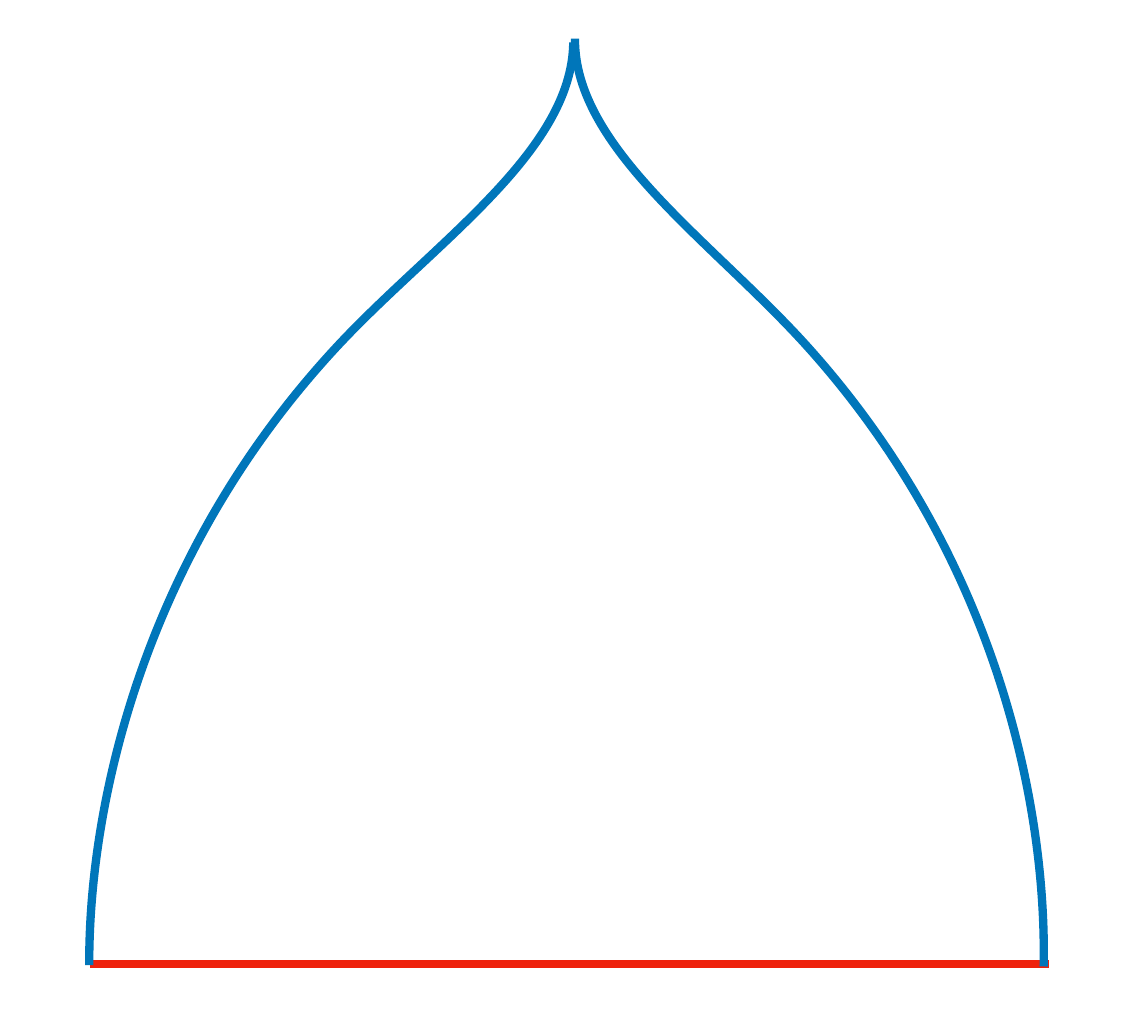} 
 \caption{Geometry underlying the AdS wave function.}
\label{fig:WFAdS}
\end{figure}

At infinite cutoff one can consider an alternative inner product by
integrating the KG current along a line at fixed $\hat{h}$:
$\Sigma' = \{h=\hat{h},\phi\}$,
\begin{equation}\label{ip2}
  \langle \hat{\psi}_{T+,\alpha'}, \hat{\psi}_{T+,\alpha} \rangle' 
  = \frac{i}{2}\int_{-\infty}^{\infty} d\phi
  \hat{\psi}_{T+}(h,\phi;\alpha') \partial_\phi\hat{\psi}_{T+}(h,\phi;\alpha) +
  \text{c.c.} \ .
\end{equation}
With
respect to this inner product the wavefunctions $\hat{\psi}_{T+}(h,\phi;\alpha)$
appear to form an orthonormal basis of a Hilbert space, labelled by the
continuous parameter $\alpha$ \cite{Held:2024rmg} (see appendix \ref{A:ga}), 
\begin{equation}
    \langle \hat\psi_{T+,\alpha'}, \psi_{T+,\alpha} \rangle'
    = \frac{1}{2\pi\lambda^2\alpha}\delta(\alpha - \alpha').
\end{equation}
However, because of the singularity at $h=\alpha^2$, this inner
product is not valid for all $h > 0$.
Moreover, the Hilbert space of finite-cutoff dS JT gravity remains an open question.
We conclude that it appears impossible to define a consistent KG inner
product for a Hilbert space of dS JT gravity.

For Lorentzian $\text{AdS}_2$ the Hilbert space of JT gravity has been studied in
\cite{Blommaert:2025bgd}. Here, a collapsing Milne universe replaces the
future trumpet of dS JT gravity, and $\alpha$ is now the size of a
geodesic circle at $\phi = 0$ (see fig.~\ref{fig:WFAdS}).  From the dS wavefunction
\eqref{tHM} one obtains by analytic continuation the AdS wavefunction
\begin{equation}
  \hat{\psi}_{\text{AdS}+} (h,\phi;\alpha) =
  \frac{1}{\lambda\sqrt{(\alpha^2-h})}\exp{(-i\lambda\phi\sqrt{\alpha^2-h})}
  \ , \quad h < \alpha^2 \ .
\end{equation}
Now $\alpha$ is the maximal size of a circle in the AdS spacetime and
the inner product \eqref{ip2} is valid for all $h > 0$ that are
  allowed in $\text{AdS}_2$. Hence, in the
case of $\text{AdS}_2$ the wavefunctions $\hat{\psi}_{\text{AdS}+} (h,\phi;\alpha)$
indeed represent a basis of an infinite dimensional Hilbert space.

The main reason for the problem to define a Hilbert space for dS
  JT gravity are the singularities at $\sqrt{h}=\alpha$ of the exact solutions of the WDW
  equation \cite{Iliesiu:2020zld,Fanaras:2021awm,Buchmuller:2024ksd}.
For some solutions, combining out- and ingoing branches into a
  real-valued function can remove these de Sitter-radius
  singularities. However, this does not work for all classes
  of solutions, and it fails in particular for the trumpet
    wavefunctions corresponding to flat operator ordering \cite{Held:2024rmg}.
  Moreover, for some trumpet wavefunctions the asymptotic behaviour at large $h$
  is an additional problem, which is relevant for trumpet wavefunctions motivated by holography. This suggests that dS JT gravity is only
  meaningful as an effective theory at intermediate length scales and
  that at short and/or large distances some modifications are needed.
Sine dilaton gravity is an example that achieves this
\cite{Blommaert:2025avl,Blommaert:2025bgd,Blommaert:2025rgw}.
Alternatively, one may add to the theory additional fields or branes
(see, for example, \cite{Friedrich:2024aad}) or singular sources
\cite{Fanaras:2021awm,Buchmuller:2024ksd}.

\section{Summary and conclusions}
\label{sec:conclusion}

%?????????????????????
%
%The definition of a ground state, a state of `minimal excitation' , is
%a subtle problem of gravity in de Sitter space.  Since forty years,
%the no-boundary proposal of Hartle and Hawking is a leading candidate
%although a number of issues remain to be settled. These include
%problems of the path integral for complex manifolds, the validity of
%saddle-point approximations, and, on the phenomenological side,
%the realisation of a sufficiently long period of inflation (for a
%discussion and references, see \cite{Maldacena:2024uhs}).
%
%In recent years new insights have been gained from studying nearly de
%Sitter space in two-dimensional Jackiw-Teitelboim gravity. In this
%model the asymptotic behaviour of the no-boundary wave function for
%large field values can be computed exactly
%\cite{Maldacena:2019cbz,Cotler:2019nbi}, which leads to a prediction
%for the probability distribution of the size of the universe.
%However, the wave function has a power-singularity at the de Sitter radius
%\cite{Iliesiu:2020zld}. Hence, it is not a solution of the WDW equation and not
%normalisable \cite{Fanaras:2021awm}.
%
%?????????????????????

A major goal of quantum cosmology is a prediction for the size of
  the universe based on quantum gravity. Given the difficulties of
  the problem, during the past years some of the conceptual questions
  have been studied in two-dimensional models of gravity. Due to its
  simplicity, Jackiw-Teitelboim gravity turned out to be particularly interesting,
  and significant progress has been made for Anti-de Sitter space
  as well as de Sitter space.

  In this paper we further pursue a recent proposal for a probability density 
  \cite{Buchmuller:2026qgp} that is based on the density matrix in
  quantum gravity \cite{Page:1986vw}. In JT gravity, after integrating
  out the dilaton, an amplitude with two boundary surfaces
  automatically takes the form of a density matrix, as a sum over
  products of trumpet wavefunctions \cite{Saad:2019lba}. These are
  transition amplitudes between a geodesic surface with vanishing
  extrinsic curvature and a second surface with Dirichlet boundary
  conditions $h$ and $\phi$ for metric and dilaton, respectively.
  In \cite{Buchmuller:2026qgp} we argued for such a form by tracing
 over the initial size of the universe which appears as a parameter of exact
 solutions of the WDW equation. It appears, however, that such an
 argument is not needed and that the form of the density matrix
 follows directly from the geometry of JT gravity, via
 analytic continuation from Euclidean AdS.

  Trumpet wavefunctions have been computed at infinite cutoff for
    $\text{EAdS}_2$ \cite{Saad:2019lba} and for $\text{dS}_2$
    \cite{Cotler:2019nbi}. At finite cutoff they have been obtained
    for $\text{EAdS}_2$ from the boundary path integral by summing up
    an infinite number of extrinsic curvature terms
    \cite{Iliesiu:2020zld}, and recently also from a bulk path integral
    \cite{Griguolo:2025kpi}. For $\text{dS}_2$, finite cutoff trumpet
    wavefunctions can be obtained by analytic continuation. The various
    results are consistent and the wavefunctions satisfy WDW equations with
    non-flat factor ordering.
    
At large $\sqrt{h}$, trumpet wavefunctions acquire prefactors which
  are monomial power-law functions of $\phi/\sqrt{h}$. The power
  appearing is related to the factor ordering in the WDW equation for
  the trumpet wavefunction constructed from the bulk path
  integral. Demanding holographic duality with the boundary curve
  Schwarzian path integral uniquely fixes the prefactor of the trumpet
  wavefunction to $(\phi/\sqrt{h})^{1/2}$.

Further, we explained that the trumpet wavefunctions always appear
  with two branches describing an expanding and a contracting dS
  universe, respectively. For each branch separately, the trumpet wavefunction
  develops a pole at $\sqrt{h}=\alpha \geq 0$. In contrast, real-valued trumpet
  wave functions built by linear superpositions of expanding and
  contracting branch can be smooth at the de Sitter radius $\sqrt{h}=\alpha$.
    
We then contrasted the trumpet wavefunctions with the no-boundary
  wavefunctions that are constructed via convolution of a trumpet
  wavefunction with a singular cap amplitude, which effectively
  describes a boundary source placed at the classical de Sitter radius
  $\sqrt{h} =\sqrt{h_c} =\lambda^{-1}$. 
  The double-trumpet amplitude, which represents a mixed-state contribution to
  the density matrix, has a semiclassical limit that agrees with the
  density matrix obtained in \cite{Fumagalli:2024msi} from a ``connected''
  wormhole geometry. It implies a prefactor $\phi/\sqrt{h}$.
  In contrast, the pure-state no-boundary contribution corresponding to a ``disconnected'' Hartle-Hawking geometry yields a prefactor
  $(\phi/\sqrt{h})^3$, and therefore a faster falloff at large $\sqrt{h}$.\\

A probability density is given by the diagonal elements of a conditional
density matrix. Following \cite{Maldacena:2019cbz}, we impose the
condition of a fixed dilaton value, $\phi = \phi_b$. $p(h|\phi_b)
= \rho(h,\phi_b;h,\phi_b)$ is then a probability density for $h$, given
this particular value of $\phi_b $. An
important consistency condition is that in the semiclassical limit
$\rho(h,\phi;h,\phi)$ is constant on a chosen classical trajectory, fixed
by the initial values $(h_0,\phi_0)$ \cite{Halliwell:1992cj,Vilenkin:1993ym}.
At large $h$ and $\phi$, in dS JT gravity a classical trajectory
satisfies $\phi/\sqrt{h} \simeq \phi_0/\sqrt{h_0}$. Hence, the
consistency condition is satisfied for any probability density that is a 
function of $\phi/\sqrt{h}$. Trumpet wave functions with holographic duals lead to a probability density
$\rho \sim \phi/\sqrt{h}$ while the no-boundary wavefunction implies
$\rho \sim (\phi/\sqrt{h})^3$. Hence, both probability densities
satisfy the consistency condition. On the other hand, the trumpet
wavefunction proposed in \cite{Held:2024rmg} leads to probability
densities $\rho \sim h^{-1}$ and $\rho \sim (\phi/h)^2$ for the connected
and disconnected part, respectively. These densities do not satisfy
the consistency condition.

The conditional probability distributions for the size of the
universe are given by $dP(h|\phi_b) \sim \rho\sqrt{h} d\sqrt{h}$.
It is remarkable that the distribution corresponding to the
double-trumpet amplitude consistent with holography ($\rho\sim \phi/\sqrt{h}$) is flat in
$\sqrt{h}$, which is consistent with inflation. All other probability
distributions fall off at large $\sqrt{h}$, disfavouring large
universes.

The trumpet wavefunctions, for all considered factor orderings, are labeled
  by the size of the geodesic circle. They are interesting candidates
  for a basis of an infinite-dimensional Hilbert space of dS JT
  gravity. Unfortunately, due to the poles at the de Sitter radii
  $\alpha$, all trumpet wavefunctions are unnormalizable, and we have
  not been able to define a Hilbert space. As a consequence, also the
  probability distribution derived from the density matrix is unnormalizable.
  However, this may even be plausible since an observer in a closed
  universe would not know how to normalize the density matrix
  \cite{Witten:2026whn}. We are then led to the conclusion that dS JT
  gravity should only be considered as an effective theory for a
  finite range of length scales, and that a complete theory requires
  additional degrees of freedom.

  \subsection*{Acknowledgments}
We thank Andreas Blommaert and Lorenzo Russo for helpful
  discussions during the DESY Theory Workshop 2026.
  AW is partially supported by the Deutsche Forschungsgemeinschaft under Germany’s Excellence Strategy - EXC 2121 “Quantum Universe” - 390833306, by the Deutsche Forschungsgemeinschaft through a German-Israeli Project Cooperation (DIP) grant “Holography and the Swampland”, and by the Deutsche Forschungsgemeinschaft through the Collaborative Research Center SFB1624  “Higher Structures, Moduli Spaces, and Integrability’'.

\appendix

\section{Group averaging inner product}
\label{A:ga}

For completeness, in this section we briefly review 
trumpet wavefunctions based on the group averaging product (for a
discussion and references, see \cite{Witten:2022xxp,Held:2024rmg}).
We follow derivations for Lorentzian $\text{dS}_2$ in
\cite{Held:2024rmg}, for Lorentzian $\text{AdS}_2$ in \cite{Blommaert:2025bgd}
and for Euclidean $\text{AdS}_2$ in \cite{Griguolo:2025kpi}.
We shall consider Lorentzian $\text{dS}_2$ throughout, and we choose the dilaton
field range $-\infty < \phi <\infty$.

As a basis in Hilbert space one can consider
momentum wavefunctions $\psi(a,k) = \int d\phi
e^{-ik\phi}\psi(a,\phi)$. With $k = -i\partial_\phi$ and $\phi =
i\partial_k$, the Hamiltonian $a\mathcal{H}_{\text{WDW}}$ (see
eq.~\eqref{wdw0}) becomes\footnote{In this section we use the
  variables $a=\sqrt{h}$ and $\phi$, and we set $\lambda = 1$.}
\begin{equation}\label{Hk}
  H = i(k\partial_a + a\partial_k) \ ,
\end{equation}
which commutes with the operator
\begin{equation}
  G = a^2 - k^2 \ .
\end{equation}  
The wavefunctions \cite{Held:2024rmg}
\begin{equation}\label{kpsib}
  \tilde{\psi}_\alpha(a,k) = 2\delta(a^2 - k^2 - \alpha^2)
\end{equation}
are eigenfunctions of both $H$ and $G$ with eigenvalues zero and
$\alpha^2$, respectively,
\begin{equation}
  H \tilde{\psi}_\alpha = 0 \ , \quad G \tilde{\psi}_\alpha = \alpha^2 \tilde{\psi}_\alpha \ .
\end{equation}
Normalizable wavefunctions $\tilde{\psi}_\alpha$ represent states of a
physical Hilbert space.\footnote{We
  restrict our discussion to the ``bounce sector'' of the Hilbert
  space. For a treatment of other sectors, see \cite{Cotler:2024xzz,Held:2024rmg}.}

The states $|a,k\rangle$ form an orthonormal basis of a pre-Hilbert space
$\mathcal{H}_0$ with inner product
\begin{equation}\label{ipH0}
  \langle a,k| a',k' \rangle = \delta(a - a')\delta(k-k') \ .
\end{equation}  
The Hamiltonian \eqref{Hk} generates boosts in $\mathcal{H}_0$,
\begin{equation}
  e^{-iHu} |a,k\rangle = |a\cosh u + k \sinh u, a \sinh u + k \cosh u
  \rangle \ .
  \end{equation}
Elements of the physical Hilbert space $\mathcal{H}$ are
``co-invariant'' states, equivalence classes that contain elements of
$\mathcal{H}_0$ as representatives. The inner product \eqref{ipH0} of $\mathcal{H}_0$
is reduced to an inner product of $\mathcal{H}$ by means of
group averaging (see \cite{Witten:2022xxp,Held:2024rmg}),
\begin{equation}
\langle a,k|\eta|a',k'\rangle = \langle a,k| \int_{-\infty}^{\infty} du 
e^{-iHu} |a',k'\rangle \ .
\end{equation}    
Here $\eta = 2\pi \delta(H)$ acts like a projection operator onto
the physical Hilbert space.
One easily verifies, as an example, that the wavefunction
$\tilde{\psi}_\alpha$ can be generated by group averaging from a state in the pre-Hilbert
space ($a^2 > k^2$),
\begin{equation}\label{ipk0}
\begin{split}
\langle a,k|\eta|\alpha,0\rangle  
&= \alpha^{-1}\delta(\sqrt{a^2-k^2}-\alpha) \\
&= 2\delta(a^2 - k^2 - \alpha^2) = \tilde{\psi}_\alpha(a,k) \ .
\end{split}
\end{equation}
Setting $k = 0$, one concludes that the states $|\alpha,0\rangle$
form an orthogonal basis of a physical Hilbert space,
\begin{equation}\label{ipga}
\langle \alpha',0|\eta|\alpha,0\rangle = \alpha^{-1}\delta(\alpha'-\alpha) \ .
\end{equation}
{Note, however, that the norm is divergent.}

In coordinate space, the wavefunctions of the basis states are given by
\begin{equation}\label{bwf}
  \begin{split}
  \psi_\alpha(a,\phi) &= \int\frac{dk}{2\pi} e^{ik\phi} \tilde{\psi}_\alpha(a,k) \\
  &= \frac{1}{\pi\sqrt{a^2 - \alpha^2}}\cos{(\phi\sqrt{a^2-\alpha^2})}
  \Theta{(a-\alpha)} \ .
\end{split}
\end{equation}
The observable $G$ acts on these wavefunctions as
\begin{equation}
  G \psi_\alpha(a,\phi) = \left(a^2 + \partial_\phi^2\right) \psi_\alpha(a,\phi)
  = \alpha^2 \psi_\alpha(a,\phi) \ .
\end{equation}
The real eigenfunctions \eqref{bwf} can be decomposed into outgoing and
incoming modes,
\begin{equation}\label{bwfpm}
  \begin{split}
  \psi_\alpha(a,\phi) &= \psi_{\alpha+}(a,\phi) + \psi_{\alpha-}(a,\phi) \ , \\ 
  \psi_{\alpha\pm}(a,\phi) &= \frac{1}{2\pi\sqrt{a^2 - \alpha^2}} e^{\mp
    i\phi\sqrt{a^2-\alpha^2}}
  \Theta{(a-\alpha)} \ ,
\end{split}
\end{equation}
for which the KG inner product reads
\begin{equation}\label{kgphi}
  \begin{split}
\langle\psi_{\alpha'\pm},\psi_{\alpha\pm}\rangle &= \mp \frac{i}{2}
\int d\phi\left(\psi_{\alpha'\pm}(a,\phi)
  \partial_\phi \psi^*_{\alpha\pm}(a,\phi) -
  \psi_{\alpha\pm}(a,\phi)\partial_\phi \psi^*_{\alpha'\pm}(a,\phi) \right)\\
&= \pm\frac{1}{2\pi\sqrt{a^2 - \alpha'^2}}
\delta(\sqrt{a^2-\alpha'^2} -\sqrt{a^2-\alpha^2}) \\
&= \pm\frac{1}{2\pi\alpha}\delta(\alpha'-\alpha) \ .
  \end{split}
\end{equation}

This suggests that the wavefunctions $\psi_{\alpha\pm}$ may form an
  orthonormal basis of a Hilbert space. However, the norm is divergent.
Moreover, the KG inner product
  should not depend on the integration path. Changing the integration
  over $\phi$ with $a$ fixed to an integration over $a$ with $\phi$
  fixed, one encounters the problem already mentioned in section~\ref{sec:hs}:
  Because of the singularity at $h=\alpha^2$ the integrand appearing in the inner product
  \begin{equation}
\langle\psi_{\alpha'\pm},\psi_{\alpha\pm}\rangle = \mp \frac{i}{2}
\int d a\left(\psi_{\alpha'\pm}(a,\phi)
  \partial_a \psi^*_{\alpha\pm}(a,\phi) -
  \psi_{\alpha\pm}(a,\phi)\partial_a \psi^*_{\alpha'\pm}(a,\phi) \right)
\end{equation}
is divergent. Here, we quickly see that the integral $\int d\sqrt{h}$ is well-defined only as long as the integrand remains finite -- that is, only as long as $a=\sqrt{h}\neq \alpha,\alpha'$. Hence, the pole in the prefactor of the out- and ingoing branches of the states eq.~\eqref{bwfpm} puts a question mark on the validity of the above integral to define a proper inner product.

In addition, there is a problem with this integral appearing when integrating towards large $\sqrt{h}$. By changing integration variables from $\sqrt{h}$ to $z\sim 1/\sqrt{h}$ this integral can be shown to be proportional to
\begin{equation}
\int \frac{d\sqrt{h}}{h} e^{i\phi\sqrt{h}} \sim  \int \frac{dz}{z} e^{i\frac{\phi}{z}}
\end{equation}
where $z\to 0$ corresponds to $\sqrt{h}\to\infty$. Hence, this supposedly equivalent way to define the KG inner product develops an essential singularity in the integrand in the limit of infinite spatial volume $\sqrt{h}$ of 2d de Sitter spacetime. Regularizing this integral requires specifying an integration contour, without which it is ill-defined. 

Both problems seem to imply that the states eq.~\eqref{bwfpm} do not form a Hilbert space with well-defined inner product for the case of JT gravity on 2d de Sitter spacetime. Moreover, we may consider both problems as signs that a quantum gravity description of global de Sitter may be difficult already in 2d.

With respect to the KG product, only
$\psi_{\alpha+}$ or $\psi_{\alpha-}$ can have positive norm, which is
inconsistent with a probability interpretation of the KG inner product. As
explained in \cite{Witten:2022xxp,Held:2024rmg}, the group averaging product
leads to equal signs for the norm of $\psi_{\alpha+}$ and $\psi_{\alpha-}$.
From eq.~\eqref{kgphi} one then obtains for the real wavefunctions
$\psi_\alpha$, $\psi_{\alpha'}$,
\begin{equation}
\langle \alpha',0|\eta|\alpha,0\rangle' = \langle \psi_{\alpha'+},\psi_{\alpha+}\rangle
- \langle \psi_{\alpha'-},\psi_{\alpha-}\rangle = \frac{2}{\alpha}\delta(\alpha-\alpha') \ ,
\end{equation}
which agrees with the group averaging inner product \eqref{ipga} up to
a factor 2. This is in contrast to the KG product which yields zero.

Similar to eq.~\eqref{ipk0}, the inner product of states with arbitrary
values of $k$ and $k'$ is given by 
\begin{equation}\label{ipkk'}
\langle a,k|\eta|a',k'\rangle  = 2\delta(a^2 - k^2 - a'^2+ k'^2)  \ .
\end{equation}
This matrix element can be written as an integral over products of basis wavefunctions,
\begin{equation}\label{kk'bb}
  \langle a,k|\eta|a',k'\rangle  = \int \alpha d\alpha\
  \tilde{\psi}_\alpha(a,k) \tilde{\psi}_\alpha(a',k') \ .
\end{equation}
Taking the Fourier transform w.r.t. $k$ and $k'$, one obtains from eq.~\eqref{ipkk'}
\begin{equation}\label{ipphik'}
  \begin{split}
    \langle a,\phi|\eta|a',\phi'\rangle  &= \int\frac{dkdk'}{(2\pi)^2} e^{i(k\phi+k'\phi')}
    \langle a,k|\eta|a',k'\rangle \\
    &= i\int\frac{dkd\tilde{k}}{(2\pi)^2} e^{i(k\phi+i\tilde{k}\phi')}
    \frac{1}{\sqrt{k^2+\tilde{k}^2}} \delta(\sqrt{a^2-a'^2} -
    \sqrt{k^2+\tilde{k}^2}) \ ,
    \quad a^2 > a'^2 \ ,\\
\end{split}
\end{equation}
where we have defined $k' = i\tilde{k}$. With
\begin{equation}
  \begin{split}
  &k = \sqrt{k^2+\tilde{k}^2} \cos \Theta\ , \quad
 \tilde{k} = \sqrt{k^2+\tilde{k}^2} \sin \Theta\ , \\
 &\frac{\phi}{\sqrt{\phi^2-\phi'^2}} =\cos \delta\ , \quad
 \frac{i\phi'}{\sqrt{\phi^2-\phi'^2}} =\sin \delta \ , \quad \phi >
 \phi' \ ,\\
&\frac{\phi}{\sqrt{\phi'^2-\phi^2}} =i\cos \delta\ , \quad
 \frac{i\phi'}{\sqrt{\phi'^2-\phi^2}} =i\sin \delta \ , \quad \phi <
 \phi' \ ,
 \end{split}
\end{equation}
this yields ($a > a'$)
\begin{equation}\label{GF}
\begin{split}
  \langle a,\phi |\eta|a',\phi'\rangle  &= \frac{i}{(2\pi)^2}
\int_0^{2\pi}d\Theta  \Big(e^{i\sqrt{a^2-a'^2}\sqrt{\phi^2-\phi'^2}
  \cos{(\Theta - \delta)}} \Theta(\phi-\phi') \\
& \hspace{3cm} + e^{-\sqrt{a^2-a'^2}\sqrt{\phi^2-\phi'^2}
  \cos{(\Theta - \delta)}} \Theta(\phi'-\phi) \Big) \ ,
\end{split}
\end{equation}
which is a sum over Bessel functions,
\begin{equation}
  \begin{split}
\langle a,\phi |\eta|a',\phi'\rangle  &= 
\frac{i}{2\pi} \Big(J_0(\sqrt{(a^2-a'^2)(\phi^2-\phi'^2)})
  \Theta(\phi-\phi') \\
& \hspace{1.2cm} + I_0(\sqrt{(a^2-a'^2)(\phi'^2-\phi^2)})
\Theta(\phi'-\phi) \Big) \ .
 \end{split}
\end{equation}
This is the transition amplitude considered in
\cite{Halliwell:1990tu,Buchmuller:2024ksd,Griguolo:2025kpi}.
%The expression \eqref{GF} is a Green function in light-cone coordinates.
The choice of the integration contour in momentum space corresponds to
a retarded Green function (see, for example,
\cite{Itzykson:1980rh,Buchmuller:2024ksd}).

%Recently, it has been shown that this expression is the Fourier transform of a
%Bessel function\footnote{Here, we have used the integral representation of $J_0$: $\int%_0^1 dx \frac{1}{\sqrt{1-x^2}}\cos(px) 
%\cosh(q \sqrt{1-x^2})=\frac{\pi}{2} J_0(\sqrt{p^2-q^2})$, see eq. (4.124) on p.~522 of~%\cite{GradshteynRyzhik2014} .} \cite{Griguolo:2025kpi}
%\begin{equation}\label{ipphiphi'}
%  \begin{split}
%    \langle a,\phi|\eta|a',\phi'\rangle  &= \int\frac{dk'}{2\pi} e^{ik'\phi'}
%    \langle a,\phi|\eta|a',k'\rangle \\
%& =\frac{-i}{2\pi} J_0(\sqrt{(\phi^2-\phi'^2)(a^2-a'^2)}) \ .
%\end{split}
%\end{equation}
%
Taking now the Fourier transform of
$\tilde{\psi}_\alpha(a,k)\tilde{\psi}_\alpha(a',k')$ in eq.~\eqref{kk'bb}
w.r.t. $k$ and $k'=i\tilde{k}$, and integrating again $k'$ along the imaginary axis,
one finds ($a > a'$)
\begin{equation}\label{phiphi'bb}
%  \begin{split}
  \langle a,\phi|\eta|a',\phi'\rangle  = \int_0^\infty \alpha d\alpha\
  \psi_\alpha(a,\phi) \bar{\psi}_\alpha(a',\phi')
   \Theta(a-\alpha)\Theta(\alpha-a')  \ .
%& + \bar{\psi}_\alpha(a,\phi) \psi_\alpha(a',\phi')
%   \Theta(a'-\alpha)\Theta(\alpha-a) \big) \ .
%\end{split}
\end{equation}
Here $\bar{\psi}_\alpha(a,\phi)$ is obtained by analytically continuing
$\psi_\alpha(a,\phi)$ from $a > \alpha$ to $\alpha > a$.

Hence, the transition amplitude between two surfaces with specified
Dirichlet boundary conditions can be expressed as an integral over
products of trumpet wavefunctions rather than an integral over
products of transition amplitudes with Dirichlet boundary conditions.
This corrects an assumption made in \cite{Buchmuller:2026qgp}.

The above structure forms the background for our definition in the
  main text of the density matrix $\rho_+(h,\phi_b;h',\phi_b)$ for
  outgoing universes in eq.~\eqref{cdm}, similar to
  eq.~\eqref{phiphi'bb} but
  with integration over $\alpha$ restricted to a finite range $[0,\kappa]$.

\bibliographystyle{utphys}
\bibliography{JT}
%JHEP.bst
\end{document}